\documentclass[prl,twocolumn,showpacs]{revtex4}
\usepackage{graphicx}
\usepackage{amsmath}

\begin{document}
\title{Single-Photon Transistor Using a F\"orster-Resonance}
\author{Daniel Tiarks}
\author{Simon Baur}
\author{Katharina Schneider}
\author{Stephan D\"urr}
\author{Gerhard Rempe}
\affiliation{Max-Planck-Institut f\"{u}r Quantenoptik, Hans-Kopfermann-Stra{\ss}e 1, 85748 Garching, Germany}


\begin{abstract}
An all-optical transistor is a device in which a gate light pulse switches the transmission of a target light pulse with a gain above unity. The gain quantifies the change of the transmitted target photon number per incoming gate photon. We study the quantum limit of one incoming gate photon and observe a gain of 20. The gate pulse is stored as a Rydberg excitation in an ultracold gas. The transmission of the subsequent target pulse is suppressed by Rydberg blockade which is enhanced by a F\"orster resonance. The detected target photons reveal in a single shot with a fidelity above 0.86 whether a Rydberg excitation was created during the gate pulse. The gain offers the possibility to distribute the transistor output to the inputs of many transistors, thus making complex computational tasks possible.
\end{abstract}

\maketitle

The gain, or equivalently the fan-out, is a central figure of merit for any physical implementation of a transistor. It describes how much a target output changes with the gate input. Gain is important for analog amplifiers, for repeaters in long-distance communication, and for digital computing. A gain of two or higher offers the perspective to distribute the output signal to the input ports of more than one transistor without attenuating the signal in each such step. Only this makes complicated computational tasks feasible.

The excessive present-day use of optical technologies in long-distance signal transmission suggests that it is desirable top apply all-optical techniques also to signal processing \cite{Miller:10, Caulfield:10}, especially because this offers perspectives to operate at high speed and low power dissipation. The cornerstone device for all-optical computing is the all-optical transistor, a device in which a gate light pulse switches the transmission of a target light pulse with a gain above unity. The fundamental low-power limit of the all-optical transistor is reached when the incoming gate pulse contains only one photon, which is interesting for a variety of applications in quantum information processing \cite{nielsen:00, Bermel:06, Chang:07}, including heralded quantum memories for quantum repeaters \cite{briegel:98}, efficient detection of optical photons \cite{Reiserer:13}, and Schr\"{o}dinger-cat states \cite{Gheri:97}.

All-optical switching with incoming gate photon numbers between a few hundred and $\sim$20 has been experimentally demonstrated in various systems, see e.g.\ Refs.\ \cite{Hwang:09, Bajcsy:09, Englund:12, Bose:12, Volz:12, Loo:12}. Even gain has been observed for 2.5 to 5 incoming photons \cite{Chen:13}. However, the single-photon regime remained elusive. In fact, it was possible to implement an all-optical switch operating at one incoming gate photon, or even fewer \cite{Baur:14}, but the gain was only $\sim$0.24. That experiment used electromagnetically induced transparency (EIT) \cite{fleischhauer:05} with Rydberg states \cite{Mohapatra:07} to store the incoming gate photon as a Rydberg excitation. The optical properties of the medium were strongly altered by Rydberg blockade, see e.g.\ Refs.\ \cite{Jaksch:00, Tong:04, Pritchard:10, gorshkov:11, Dudin:12, Peyronel:12, Parigi:12}, resulting in a suppression of the transmission of a subsequent target pulse. A large principal quantum number of $n=100$ was used for the gate and target pulse to achieve good blockade. Different polarizations for gate and target light were used to reduce undesired retrieval of the gate photons by target light. But several problems posted severe obstacles for gain in Ref.\ \cite{Baur:14}. First, the remaining undesired retrieval of the gate excitations by target light deteriorated the target suppression for long target pulses, thus making long target pulses useless. Second, self blockade of the target light set an upper bound on the transmitted target signal power. Third, dephasing reduced the peak transmission on the EIT resonance and reduced the transmitted target signal power.

Here we experimentally demonstrate an all-optical transistor with one incoming gate photon on average and a gain of 20(1) per incoming photon. We use principal quantum numbers $n_g= 69$ and $n_t= n_g-2$ for the gate and target pulse, respectively. The fact that the quantum numbers differ results in drastically better suppression of undesired retrieval. As a result, the length of the target pulse can be increased by roughly two orders of magnitude without a strong deterioration of the extinction. The fact that the principal quantum numbers are lower reduces self blockade and dephasing. In addition, the lower quantum numbers increase the population lifetime of the Rydberg state at the densities of our experiment because inelastic collision rates decrease. A F\"orster resonance \cite{Walker:08} for the chosen principal quantum numbers achieves good blockade between gate and target pulse, despite the fact that the principal quantum numbers are not very large. We experimentally study how extinction and gain profit from the F\"orster resonance. We observe that the transmitted signal light has a bimodal photon number distribution if the gate pulse is applied. This is because the storage efficiency is below unity. Based on this bimodal distribution, we determine whether a Rydberg excitation was created during the gate pulse with a single-shot fidelity above 0.86. Unlike Ref.\ \cite{Chen:13}, the gate and target pulse in our experiment operate at the same wavelength, which is important for many applications.

\begin{figure}[!t]
\centering
\includegraphics[scale=0.8]{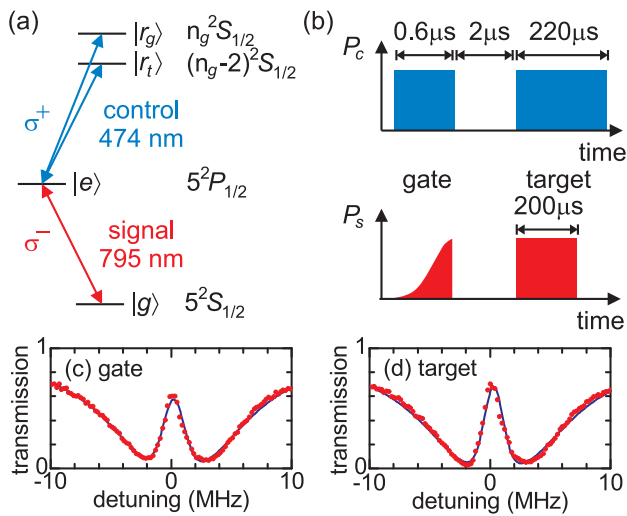}
\caption{\label{fig-scheme}
(color online) (a) Level scheme. Gate and target pulse each consist of signal light and control light for EIT. Both pulses use the same signal transition, but the control light operates at different frequencies, thus reaching different Rydberg states $|r_g\rangle$ and $|r_t\rangle$. The hyperfine quantum numbers are $F=1,m_F=-1$ and $F=2,m_F=-2$ for states $|g\rangle$ and $|e\rangle$, respectively, whereas both Rydberg states have $m_J=1/2$ and $m_I=-3/2$. The fact that the principal quantum numbers $n_g$ and $n_t= n_g-2$ differ suppresses undesired retrieval of stored gate excitations by target control light much more efficiently than the polarization scheme of Ref. \cite{Baur:14}. In addition, both signal light pulses profit from a large electric-dipole matrix element. We typically operate at $n_g=69$. (b) Input power timing scheme, not to scale (see text). (c,d) EIT spectra (see text).
}
\end{figure}

The experimental setup was described and characterized in detail in Ref.\ \cite{Baur:14}. In brief, an ultracold gas of $^{87}$Rb with atom number $N\sim 1.5\times10^5$ and temperature $T=0.33$ $\mu$K is held in an optical dipole trap with trapping frequencies of $(\omega_x,\omega_y,\omega_z)/2\pi= (136,37,37)$ Hz. It is illuminated by a signal light beam at a wavelength of $\lambda_s= 795$ nm which propagates along the horizontal $z$ axis. This beam is used for gate and target pulses. Two control light beams originate from two different lasers. One is used for the gate pulse, the other for the target pulse. The gate control beam counterpropagates the signal beam, and has a wavelength of $\lambda_{c,g}= 474$ nm. The target control beam copropagates with the signal beam, and is several tens of gigahertz red detuned from the gate control light. The beam waists ($1/e^2$ radii of intensity) are $(w_s,w_{c,g},w_{c,t}) = (8,21,12)$ $\mu$m. The blockade radius is estimated to be $r_b=16\ \mu$m \cite{note:blockade-radius}. The control beams have powers of $(P_{c,g},P_{c,t})= (17,10)$ mW. A magnetic field of 1.1 G is applied along the $z$ axis. The probability for collecting and detecting a transmitted signal photon is $\eta_\text{det}= 0.24$.

Parts (a) and (b) of Fig.\ \ref{fig-scheme} show the atomic level scheme and the timing sequence, respectively. Gate and target pulse each consist of signal light and control light for EIT. The gate control light is switched off while a large part of the gate signal light is inside the medium, thus storing the gate signal light in the form of a Rydberg excitation. If a gate excitation was stored, Rydberg blockade will suppress the transmission of the subsequent target signal pulse. In the absence of a gate pulse, however, target signal light experiences a high transmission because of EIT. This gate-target pulse sequence is repeated with a cycle repetition time of $t_\text{cyc}=1$ ms. After $\sim$100 gate-target cycles, we prepare a new atomic sample.

Figure \ref{fig-scheme}(c) shows the dependence of the signal transmission on the signal detuning for the gate pulse but with a gate pulse durations of 200 $\mu$s and 220 $\mu$s for signal and control light, respectively. Figure \ref{fig-scheme}(d) shows the same for the target pulse if the gate pulse is omitted. Fits of the simple, empiric model of Ref.\ \cite{Baur:14} to each data set yield EIT linewidths of $\Delta_T\sim 2\pi\times1.9$ MHz (full width at half maximum) and optical depths of $OD\sim 5$. For comparison, we use the parameters of the atomic cloud to estimate the transmission averaged over the transverse beam profile $\langle T\rangle$. Equating this with $\langle T\rangle = e^{-OD}$ yields $OD\sim8$, which agrees fairly well with the above best-fit value.

\begin{figure}[!t]
\centering
\includegraphics[scale=1]{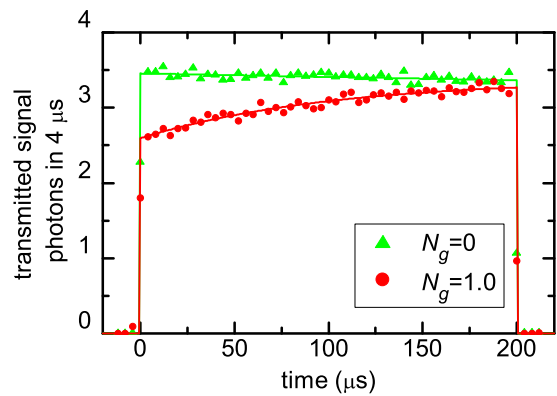}
\caption{\label{fig-trace}
(color online) Single-photon transistor. The number of transmitted signal photons is shown for $N_g=1.0$ incoming signal gate photons (red circles). For reference, the same number is shown in the absence of target signal light $N_g=0$ (green triangles). The lines show exponential fits multiplied by a step function. The ratio of the areas under the two data sets yields an extinction of $\epsilon= 0.89(1)$. The gain is $G= 20(1)$. This is far above unity, thus demonstrating a single-photon transistor.
}
\end{figure}

Figure \ref{fig-trace} shows experimental results. The number of transmitted target signal photons is shown for $N_g= 1.0$ incoming gate signal photons (red circles). The area under the curve reveals the number of transmitted target signal photons $N_\text{trans}$. A measurement with $N_g= 0$ (green triangles) yields a corresponding reference value $N_\text{trans, ref}$. The extinction
\begin{equation}
\epsilon
= \frac{N_\text{trans}}{N_\text{trans, ref}}
\end{equation}
quantifies how well the gate pulse suppresses the target pulse. The gain
\begin{equation}
G
= \frac{\Delta N_\text{trans}}{N_g}
= \frac{|N_\text{trans, ref}-N_\text{trans}|}{N_g}
\end{equation}
quantifies how many input ports of identically constructed transistors could be driven. The observed value of $G= 20(1)$ at $N_g= 1.0$ is far above unity, thus clearly demonstrating the realization of a single-photon transistor.

\begin{figure}[!t]
\centering
\includegraphics[scale=1]{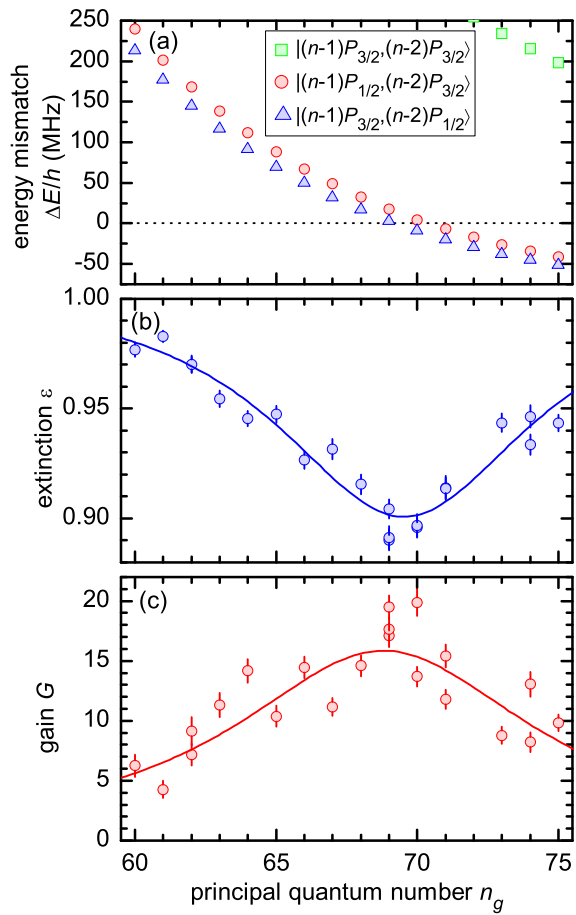}
\caption{\label{fig-Foerster}
(color online) Effect of the F\"orster resonance on the single-photon transistor. (a) Theoretical estimates of the energy mismatch $\Delta E$ in Rb at infinite interatomic distance. Some states from the fine-structure manifold of the state $|(n-1)P,(n-2)P\rangle$ lie close to the state $|nS_{1/2},(n-2)S_{1/2}\rangle$. Two energy mismatches cross zero near $n=70$, thus creating the F\"orster resonance used in our experiment. The measured values of the extinction (b) and the gain (c) clearly profit from the F\"orster resonance. The target pulse is operated at $n_t= n_g-2$. The lines show Lorentzian fits to guide the eye.
}
\end{figure}

For long target pulse duration, the transmitted target photon number approaches the reference, as seen in Fig.\ \ref{fig-trace}. We divide the data by the reference, and obtain a $1/e$ time of $\tau= 0.10(1)$ ms from an exponential fit. This is not far from the excited-state lifetime due to radiative decay at room temperature of 0.14 ms \cite{Saffman:05}, showing that undesired retrieval of gate excitations and inelastic collisions have only a small effect. The transmission of the reference target pulse is $T_0= 0.49(1)$ at the EIT resonance. This improvement by a factor of $\sim$2.5 with respect to Ref.\ \cite{Baur:14} is mostly due to the reduced principal quantum number, which reduces self blockade and dephasing. Experimentally varying $n_g$ in the range between 60 and 75, we find that $T_0$ depends approximately linearly on $n_g$ with a slope of $\Delta T_0/\Delta n_g\sim -0.01$.

The F\"orster resonance used in our experiment is caused by the fact that the energy mismatches between certain atom-pair states are close to zero near $n=70$. We calculate these energy mismatches from literature values \cite{Li:03} for the Rb quantum defects. Results of this calculation are shown in Fig.\ \ref{fig-Foerster}(a). Parts (b) and (c) show that the performance of the single-photon transistor profits from the F\"orster resonance. Extinction and gain both show a clear resonance.

Because of the gain, storing a gate excitation has a drastic effect on the transmitted light. This holds not only for the mean value $N_\text{trans}$ of the transmitted photon number but also for the probability distribution of the transmitted photon number. We measure the histogram for the number of detector clicks $N_c$ registered during a 30 $\mu$s long target signal pulse. These data are recorded with $N\sim2.4\times10^5$, $T=0.27$ $\mu$K, $(P_{c,g},P_{c,t})= (35,22)$ mW, a dark time of 0.15 $\mu$s between gate and target pulse, a target control pulse duration of 100 $\mu$s, and $t_\text{cyc}=0.7$ ms.

\begin{figure}[!t]
\centering
\includegraphics[scale=1]{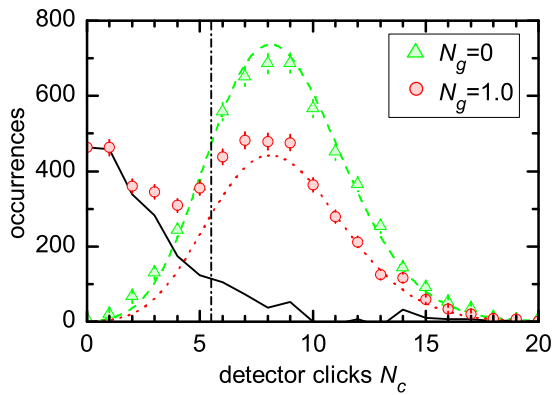}
\caption{\label{fig-bimodal}
(color online) Bimodal distribution. The histogram for the number of detector clicks $N_c$ during the target pulse is shown. The reference pulse with $N_g= 0$ (green triangles) has mean value $\langle N_c\rangle_\text{ref}= 8.62(4)$ and variance 9.87. It is well approximated by the Poisson distribution with this mean value (green dashed line). The distribution for $N_g= 1.0$ (red circles) is bimodal. Its peak near $N_c=8$ is caused by events in which zero gate excitations were stored. This peak is identical to the green line except for an overall factor $p_0$, expressing the probability for storing zero excitations. A fit (red dotted line) to the data with $N_c>9.5$ yields the best-fit value $p_0=0.60$. The black solid line shows the difference between the red data and the red fit curve. It represents the histogram if a Rydberg excitation was stored. From the value of $N_c$ measured in a single shot, one can infer whether a Rydberg excitation was stored. Setting the discrimination threshold to $N_\text{thr}=5.5$ (dash-dotted vertical line) yields a fidelity for this inference of $F=0.86$ (see text).
}
\end{figure}

Figure \ref{fig-bimodal} shows this histogram. The data for $N_g=1.0$ show a bimodal structure with a clearly visible minimum between the peaks. Obviously, the peak near $N_c=8$ detector clicks is expected to be identical to the reference distribution with $N_g=0$, but with the total number of events reduced by an overall factor $p_0$, which is the probability that zero Rydberg excitations are stored during the gate pulse. A fit (red dotted line) yields $p_0= 0.60$. Subtracting this fit from the data for $N_g=1.0$ yields the black solid line.

The red dotted line and the black solid line have well-separated peaks. Hence, the value of $N_c$ obtained in a single experimental shot reveals whether the number of Rydberg excitations $N_\text{Ryd}$ stored during the gate pulse was zero or nonzero. We set a threshold $N_\text{thr}$ and assign $N_\text{Ryd}=0$ if $N_c> N_\text{thr}$ and $N_\text{Ryd}\neq0$ otherwise. Let $c_0$ and $c_1$ denote the probability that this assignment is correct if the initial state were ideally prepared with $N_\text{Ryd}=0$ and $N_\text{Ryd}\neq0$, respectively. To define the fidelity for estimating whether a Rydberg excitation was stored, we follow the conservative definition of Ref.\ \cite{Bochmann:10} that the fidelity $F$ is the minimum of $c_0$ and $c_1$. From Fig.\ \ref{fig-bimodal} we find that the choice $N_\text{thr}=5.5$ maximizes the fidelity, yielding $F=0.86$.

Note that the results for $p_0$ and $F$ depend on the choice of the cutoff value $N_\text{cut}$ of $N_c$ for fitting $p_0$. In particular, for $N_\text{cut}\ll 9.5$ the black solid curve contributes noticeably to the red data, causing the fit to overestimate $p_0$ and $F$, whereas for $N_\text{cut}\gg 9.5$ the fit must infer the peak height only from data far out on the wings of the distribution which is prone to produce incorrect results. A detailed analysis shows that fits with $7.5\leq N_\text{cut}\leq 12.5$ produce reliable values, with $p_0$ varying between 0.60 and 0.64 and with $F$ between 0.86 and 0.88. We choose to quote $F=0.86$ as a conservative estimate.

The value of $p_0$ determined here can be used to estimate the storage efficiency $\eta_s= N_s/N_g$, where $N_s$ is the number of stored excitations. On one hand, the solid bound $N_s\geq 1-p_0$ is reached if the probability of storing more than one excitation is neglected. Using $p_0= 0.64$, we obtain $0.36\leq \eta_s$. On the other hand, as self blockade of the gate pulse is not very pronounced, one could approximate the number of stored excitations as Poissonian, so that $p_0=\exp(-N_s)$ and $p_0=0.60$ would yield the less conservative estimate $\eta_s\sim 0.51$.

In addition to the perspectives already discussed in the introduction, Fig.\ \ref{fig-bimodal} shows that our system offers an efficient method for the nondestructive detection of a Rydberg excitation in a single experimental shot with a fidelity above $F=0.86$. In the future, this method could be used for various purposes, such as to monitor the spatial and temporal dynamics of a single Rydberg excitation \cite{Muelken:07}. A recent experiment demonstrated nondestructive imaging capabilities for Rydberg atoms but was unable to reach sufficient sensitivity to detect a single Rydberg excitation in a single experimental shot \cite{Guenter:13}. Single-shot data acquisition capabilities of single excitations offer the possibility to record, on the one hand, full probability distributions and correlation functions and, on the other hand, real-time trajectories of individual excitations. This contains much more information than mean values obtained from averaging over many shots or many excitations.

A related experiment was simultaneously performed at the University of Stuttgart \cite{Gorniaczyk:1404.2876}.

This work was supported by the DFG via NIM and via SFB 631.


\begin{thebibliography}{10}
\bibitem{note:blockade-radius}
Using quantum defect theory and assuming that the internuclear axis lies along the $z$ axis, we calculate a van-der-Waals coefficient of $C_6= 2.3\times10^{23}$ atomic units for $n_g= 69$ and $n_t= n_g-2$. We use $r_b= (|C_6|/\hbar\Delta_T)^{1/6}$ from Ref.\ \cite{Pritchard:10} to calculate $r_b= 16\ \mu$m. $r_b$ is so large that the $C_6$ approximation works well. Going beyond it causes a correction of less than 10\% to $r_b$. Nevertheless, the single-photon transistor profits from the nearby state because $|C_6|$ is strongly increased. For comparison, for $n_g=n_t=69$, one finds only $C_6= -0.05\times10^{23}$ atomic units.
\end{thebibliography}
\end{document}